\documentclass[11pt]{article}
\pdfoutput=1
\usepackage{jcappub}
\usepackage{amsmath,amssymb,amsbsy}
\usepackage{mathtools}
\usepackage{mathpazo}
\DeclarePairedDelimiter\abs{\lvert}{\rvert}%
\DeclarePairedDelimiter\norm{\lVert}{\rVert}%
\makeatletter
\let\oldabs\abs
\def\abs{\@ifstar{\oldabs}{\oldabs*}}
\let\oldnorm\norm
\def\norm{\@ifstar{\oldnorm}{\oldnorm*}}
\makeatother

\def\beq{\begin{eqnarray}}
\def\eeq{\end{eqnarray}}

\def\l({\left(}
\def\r){\right)}

\begin{document}


\title{An effective field theory of holographic dark energy}

\author{Chunshan Lin\,,\qquad }

\affiliation{Faculty of Physics, Astronomy and Applied Computer Science, Jagiellonian University, 30-348 Krakow, Poland}

\abstract{
A general covariant local field theory of the holographic dark energy model is presented. It turns out the low energy effective theory of the holographic dark energy is the massive gravity theory whose graviton has 3 polarisations, including one scalar mode and two tensor modes. The Compton wavelength is the size of the future event horizon of the universe. 
The UV-IR correspondence  in the holographic dark energy model stems from the scalar graviton's strong coupling at the energy scale that marks the breaking down of the effective field theory.
}

\maketitle

\section{Introduction}
The holographic dark energy model \cite{Cohen:1998zx}\cite{Li:2004rb} is based on the simple idea that the vacuum energy density arising from the quantum fluctuation of the UV-cut-off quantum field theory should relate to the boundary surface of a system in the way 
\beq\label{rholambda}
\rho_\Lambda\sim M_p^2L^{-2},
\eeq
where $M_p^2=8\pi G$ is the reduced Planck mass, $L$ is the size of a system (thus $L^{2}$ is essentially the area of the boundary surface). This relation between the energy density in a bulk and the area of its space-time boundary surface is rooted in the holographic principle \cite{tHooft:1993dmi}. The argument that leads to the above relation is the following. The zero-point energy diverges quartically and thus the energy density scales as $\rho_\Lambda\sim\Lambda^4$ if we simply cut-off the divergence at the UV scale $\Lambda$ \footnote{Noted that there are some works \cite{Akhmedov:2002ts}\cite{Ossola:2003ku}\cite{Donoghue:2020hoh} that challenge this perspective.}.  However, this simple scaling is violated when the total energy in the system with size $L$, i.e. $L^3\Lambda^4$, approaches to the mass of the black hole in the same size $LM_p^2$. In fact, the total energy must be bounded by the mass of the black hole from above, as the effective quantum field theory breaks down at the Schwarzschild radius scale.  This energy bound is stronger than the Bekenstein entropy bound \cite{bekenstein1} which inspired the proposal of the holographic principle \cite{tHooft:1993dmi}. Nowadays it is widely believed that the holographic principle is one of the most important cornerstones of quantum gravity.

 Assuming that the energy bound is saturated on the cosmological background, one obtains the important UV-IR correspondence \beq\label{uvir}
 \Lambda\sim\sqrt{M_p/L}
 \eeq
  and the vacuum energy $\rho_\Lambda$ comparable to our current critical energy density.  The related idea was discussed in Ref. \cite{Horava:2000tb}\cite{thomas} with the Hubble radius as the IR cut-off. However, as pointed out in Ref. \cite{Hsu:2004ri},  the resultant equation of state is greater than $-1/3$ and thus the vacuum energy fails to accelerate the cosmic expansion, if we simply take the Hubble radius $L=1/H$ as the IR cut-off. Later in the same year, by adopting the future event horizon as the IR cut-off, an accelerated expanding solution in the  Friedman equation was eventually obtained in  Ref.\cite{Li:2004rb}. Interestingly, the cosmic coincidence problem can also be resolved by inflation in this scenario, provided the minimal number of inflationary e-foldings. Since then, the holographic dark energy has drawn a lot of attention and has been widely studied, see Ref. \cite{Wang:2016och} for a comprehensive review on the topic. In passing, I shall mention that it was pointed out very recently that the holographic dark energy model may alleviate the Hubble tension problem \cite{Dai:2020rfo}. 

Given the phenomenological success, however, one of key pieces for our holographic jigsaw puzzle is still missing. Namely we do not know how to write down the general covariant action for the holographic dark energy model. An attempt was made in the unpublished work \cite{Li:2012xf} in which a mini superspace action was given. Nevertheless, due to the absence of the general covariant action, one may easily spot several conceptual problems. For instance, the evolution of our universe in the past is dependent on the one in the future. This apparent causality violation may not imply the pathology of the model, but rather that the  low energy effective field theory is missing. This causality violation may be partially addressed at the cosmological background evolution level, given the mini superspace action \cite{Li:2012xf}. However, it is still not quite clear whether the local physics violates the causality, as the perturbation theory is also missing\footnote{A preliminary analysis on the perturbative stability was conducted in Ref. \cite{Li:2008zq}}. 

In the current work, I aim at finding this important missing piece for our holographic jigsaw puzzle. As we will see in the remain of this paper, the low energy effective field theory (EFT) for the holographic dark energy model is actually a massive gravity theory. A graviton has three polarisations, namely one scalar mode, and two tensor modes. The UV-IR correspondence eq. (\ref{uvir})  stems from the strong coupling of the scalar mode above the energy scale where our effective field theory breaks down. One may get puzzled at this point as the Poincare symmetry in 4 dimensional space-time implies a massive spin-2 particle has 5 polarisations (including helicity 0, $\pm1$, $\pm2$). However, the massive gravity theory that we are going to engage is different from the conventional one \cite{fp}\cite{deRham:2010kj}, as the global Lorentz invariance is broken on the cosmological background and thus a graviton does not necessary to be well equipped with  5 polarisations \cite{Dubovsky:2004sg}\cite{Comelli:2014xga}. See Ref. \cite{Hinterbichler:2011tt}\cite{deRham:2014zqa} for two comprehensive reviews on the topic of massive gravity.

This paper is organised in the following. I will start from the mini superspace action in the section \ref{sec2}, and then write down the covariant action by adopting the Stueckelberg trick. Upon writing down the general covariant action, an Ostrogradsky ghost is spotted in the decoupling limit. The ghost is eliminated in the section \ref{sec3}, in which the linear perturbation and Hamiltonian structure are analyzed in detail. The conclusive remarks and outlooks are given in the section \ref{sec4}.

\section{From mini superspace to general covariance}\label{sec2}

We take the flat FLRW ansatz $ds^2=-N^2dt^2+a^2d\textbf{x}^2$ and the following mini superspace action \cite{Li:2012xf} as our starting point, 
\beq
S=\frac{M_p^2}{2}\int dt\left[\sqrt{-g}\left(\mathcal{R}-\frac{2c}{a^2L^2}\right)-\lambda\left(\dot{L}+\frac{N}{a}\right)\right]+S_m,
\eeq
where $M_p^2\equiv 8\pi G$ is the reduced Planck mass, $\mathcal{R}$ is the 4 dimensional Ricci scalar, and $L$ is the variable with length dimension and subject to the constraint equation enforced by the Lagrangian multiplier $\lambda$, i.e. $\dot{L}=-N/a$. We may integrate this constraint equation from the infinite past $-\infty$ to nowadays, namely 
\beq
L=\int_{-\infty}^{t}\frac{-Ndt'}{a(t')}+L(-\infty),
\eeq
where $L(-\infty)$ is the initial condition in the infinite past, which requires the input from new physics (such as quantum gravity) to fully determine its value, in light of the Hawking-Penrose's singularity theorem  \cite{hawking-penrose}.  Therefore $L(-\infty)$ remains unknown to me due to my ignorance and obtuseness. On the other hand, one may try to take the integration from the other side, namely 
\beq\label{feh}
L=\int^{\infty}_{t}\frac{Ndt'}{a(t')}+L(+\infty),
\eeq
given the asymptotic solution in the infinite future $t=+\infty$. It turns out $L(+\infty)=0$ for the asymptotic solution derived in Ref. \cite{Li:2012xf} and later is confirmed by the numerical computation in Ref. \cite{Li:2012fj}. Therefore, $R_h\equiv aL$ is exactly the size of the future event horizon\footnote{Mathematically a more general solution to $\dot{L}=-N/a$ is 
\[L=\int_{t_0}^{t}\frac{-Ndt'}{a(t')}+L(t_0).\]
We may input the value of $L$ at the moment $t_0$ in the past by hand, and then treat it as the initial condition of the equations of motion, which is equivalent to introducing an assumption about the UV completion of the theory.}. It follows the energy density of the dark sector
\beq\label{rhodark}
\rho_{\text{dark}}=M_p^2\left(\frac{c}{a^2L^2}+\frac{\lambda}{2a^4}\right),
\eeq
where the first term is the holographic term, and the second term is the dark radiation as it scales as $a^{-4}$. In passing, I shall mention that in some cases, for instance, a cyclic universe realised by means of some exotic matter content, the future event horizon may not exist as the integral in the eq. (\ref{feh}) may not converge to a finite and definite value. The effective field theory of the holographic dark energy, which will be developed in the remains of the this paper, does not apply to these cases. Therefore, we will only focus on the universe models whose future event horizon has finite, definite and non-vanishing size. 

The general covariance can be recovered by Stueckelberging the mini superspace action \cite{ArkaniHamed:2002sp}. We may follow the following dictionary,
\beq\label{dic}
\frac{1}{a^2}\to\frac{1}{3}g^{ij}\delta_{ij}\to g^{\mu\nu}\partial_\mu\phi^a\partial_\nu\phi^b\delta_{ab}, &&\qquad\text{where}\qquad \langle\phi^a\rangle=\frac{x^i\delta_{ia}}{\sqrt{3}},\nonumber\\
L\to\varphi(t,\textbf{x}), \nonumber\\
-\frac{\dot{L}^2}{N^2}\to g^{\mu\nu}\partial_\mu\varphi\partial_\nu\varphi,&&\qquad\text{where}\qquad  \langle\varphi\rangle=\varphi(t),
\eeq
where $\delta_{ab}$ is the metric of the Stueckelberg scalar field space,  $\delta_{ia}$ is the pullback mapping between the physical space-time and the scalar field space, and $i,j,k...$  are adopted as 3 dimensional spatial coordinate indices, while $a,b,c...$ are adopted as the field space indices. These 4 Stueckelberg fields are not yet canonically normalized and thus they are of the length dimension. The field space is flat, and respects the $SO(3)$ rotational invariance. In passing, I shall mention that the global Lorentz invariance is broken in this scalar configuration. Nevertheless, one should not be worried about this Lorentz-violating vacuum expectation values (VEVs) of scalars as the Lorentz invariance is broken anyway on the cosmological background. For instance, CMB is not invariant under a Lorentz boost. 

Before writing down the general covariant action, which is pretty easy and straightforward at this point, I shall remind readers the hierarchical structure of the theory. We have the Planck scale as the genetic fundamental scale in the first place, and a secondary fundamental scale for our EFT of the dark  sector, which is expected to be generated  via the non-trivial VEV of the time-like Stueckelberg field $\varphi$, namely $\Lambda\sim\sqrt{M_p/\varphi}$. The low energy effective field theory that I begin with is the following, 
\beq\label{action1}
S=\int d^4x\sqrt{-g}\left\{\frac{1}{2}M_p^2\mathcal{ R}-M_p^2\varphi^{-2}\left[\left(c+\lambda\right)\cdot\partial^\mu\phi^a\partial_\mu\phi^b\delta_{ab}+\lambda\partial^\mu\varphi\partial_\mu\varphi\right]\right\}+S_m,
\eeq
where $c$ is a constant,  $\lambda$ is a Lagrangian multiplier, and $S_m$ is the action of the matter sector which minimally couples to gravity. The theory respects  the $SO(3)$ rotational invariance, $Z_2$ symmetry, and the rescaling invariance 
\beq\label{globalsym}
\phi^a\to \ell\cdot \phi^a, \qquad \varphi\to \ell\cdot \varphi,
\eeq
where $\ell$ is a constant. 
The Lagrangian multiplier $\lambda$ generates a constraint between the temporal Stueckelberg field and the spatial Stueckelberg fields. A similar example of this sort can be found in the generalized unimodular gravity \cite{Barvinsky:2020sxl}. The hierarchical structure of the theory is manifest at the action level. One would expect that the hierarchy disappears if we trace the cosmic evolution all the way backward in time to the infinite past, and it was generated later as the universe expands. The physical significance of the energy scale $\Lambda\sim\sqrt{M_p/\varphi}$ will be shown in the next section. 

I adopt the FLRW ansatz, and the four Stueckelberg scalar fields take the space-time VEVs in the eq. (\ref{dic}). The background equations of motion read,
\beq\label{friedmann}
3M_p^2H^2&=&\frac{c}{a^2\varphi^2}+\frac{\lambda}{2a^4}+\rho_m,\nonumber\\
-M_p^2\dot{H}&=&\frac{c}{3a^2\varphi^2}+\frac{\lambda}{3a^4}+\frac{1}{2}\left(\rho_m+p_m\right),
\eeq
and 
\beq\label{bgcons}
\dot{\varphi}=-\frac{1}{a},\qquad\qquad
\dot{\lambda}=\frac{-4ca}{\varphi^3},
\eeq
where $\rho_m$ and $p_m$ are the energy density and pressure of the matter sector, the lapse has been absorbed into the redefinition of time, namely $Ndt\to dt$, and I have rescaled the Lagrangian multiplier $\lambda\to\lambda\cdot\frac{\varphi^{2}}{4a^2M_p^2}$ for convenience. All equations of motion presented in Ref. \cite{Li:2012xf} have been reproduced\footnote{ The constraint equation associated with the Lagrangian multiplier $\lambda$ has two solutions $\dot{\varphi}=\pm1/a$, which are equivalent due to the $Z_2$ symmetry of the theory. Without losing generality,  we adopt the solution $\dot{\varphi}=-1/a$.}. Some detailed analyses about this set of equations, including the analytical one and the numerical one, have already been conducted in Ref.  \cite{Li:2012xf}  and Ref. \cite{Li:2012fj}, and I shall not repeat it in my current work. 

Before moving on to study the local dynamics, I shall briefly comment on the causality. The $apparent$ causality violation stems from our ignorance of the UV completion of the theory, which compels us to integrate the differential equation (\ref{bgcons}) backward from $t=\infty$ to the moment of interest, as the asymptotic value of $\varphi$ at the infinite future is known to be $\varphi(\infty)=0$ \cite{Li:2012xf}\cite{Li:2012fj}. Assuming we had enough knowledge about the quantum gravity and we were able to specify the initial condition at $t=-\infty$, we might just integrate the differential equation from $t=-\infty$ to the moment of interest, as we normally do, and there is no apparent causality violation in this setup. The result obtained from this setup should be equivalent to the one integrated backward from $t=\infty$. In other words, the $apparent$ causality violation appears as an artifactual and mathematical treatment for our equations of motion, rather than a real phenomenon.

It is quite remarkable to see that the holographic dark energy, which is $seemingly$ non local and causality-violating, actually originates from a simple and well defined local field theory. The trick of the game is the spontaneous symmetry breaking, namely that  I start from the general covariant theory, then 4 scalar fields take the non-trivial space-time VEVs and break the spatial and temporal diffeomorphism invariance around this vacuum solution. According the Goldstone theorem, there is a massless boson for each generator of the symmetry that is broken. I shall start to introduce the Goldstone excitations around this asymmetric state,
\beq
\phi^a=\frac{1}{\sqrt{3}}\left(x^i\delta_i^a+\pi^a\right),\qquad\qquad\varphi=\varphi(t)+\pi^0,
\eeq
where $\pi^a$ is the space-like Nambu-Goldstone boson, and $\pi^0$ is the time-like one. 
Due to the $SO(3)$ rotational symmetry, we can decompose the helicity $\pi^a=\delta^i_a\partial_i\pi+\hat{\pi}^a$, where $\pi$ is the longitudinal mode, and $\hat{\pi}^a$ are two transverse modes satisfying the transverse condition $\partial_i\hat{\pi}^i=0$. 

In a gauge field theory, the Goldstone pions decouple from the gauge bosons at the scale much shorter than the Compton wavelength. The physics becomes very transparent in the decoupling limit. A similar decoupling limit can also be adopted, where the Goldstone bosons decouple from the graviton, which is regarded as a gauge boson in this case. The limit is defined by setting $M_p^2\to\infty$ while keeping  $\Lambda=\sqrt{M_p/\varphi}$  fixed (I neglect the time dependence of $\varphi$ for the time being for the schematic analysis). 
It turns out this decoupling limit is a very illuminating perspective allowing us to take a quick peek at the microscopic dynamics of the dark sector. Perturbatively expanding the action eq. (\ref{action1}) up to quadratic order in the Goldstone excitations, I get the following Goldstone action in the decoupling limit,
\beq\label{goldstone1}
S_\pi\simeq\Lambda^4\int\frac{\lambda}{4}\left(\dot{\pi}^0\dot{\pi}^0-\partial_i\pi^0\partial_i\pi^0\right)+\frac{4c+\lambda}{12}\left(\partial_i\dot{\pi}\partial_i\dot{\pi}-\partial^2\pi\partial^2\pi\right)+\frac{1}{6}\delta\lambda\left(3\dot{\pi}^0-\partial^2\pi\right),
\eeq
where $\partial^2\equiv\partial_i\partial_j\delta^{ij}$.
Seemingly so far so good, but unfortunately there is a pitfall lies in the Goldstone action. Taking variation w.r.t $\delta\lambda$ we get the constraint $\partial^2\pi=3\dot{\pi}^0$. Inserting it back into the eq. (\ref{goldstone1}), we get a higher order temporal derivative term \footnote{The operator $1/\partial^2$ should be understood in the momentum space.}
\beq
S_\pi\supset \frac{12c+3\lambda}{4} \int \ddot{\pi}^0\frac{1}{\partial^2}\ddot{\pi}^0+...
\eeq
This term yields to a fourth order equation of motion, which requires 4 initial conditions to fully determine its evolution. It implies there are 2 scalar degrees of freedom in the system, instead of one. One of them is actually the infamous Ostrogradsky ghost \cite{ghost1}\cite{ghost2}. The ghost instability spoils the  validity of the low energy effective field theory, and renders the theory inconsistent. I will devote the next section to kill the ghost. 

\section{ghost elimination and  perturbation analysis}\label{sec3}

\subsection{Stueckelberg fields and gauge transformation}
I define the linear perturbation of our FLRW metric as follows, 
\beq\label{perturbation}
g_{00}&=&-N(t)^2\left(1+2\alpha\right),\nonumber\\
g_{0i}&=&N(t)a(t)\left(\partial_i\beta+S_i\right),\nonumber\\
g_{ij}&=&a(t)^2\left[\delta_{ij}+2\psi\delta_{ij}+\partial_i\partial_jE+\frac{1}{2}\left(\partial_iF_j+\partial_jF_i\right)+\gamma_{ij}\right].
\eeq
where $\alpha, \beta,\psi$ and $E$ are the scalar perturbations, $S_i$ and $F_i$ are the vector perturbations satisfying the transverse condition $\partial_iS_i=\partial_iF_i=0$, and $\gamma_{ij}$ is the tensor perturbation satisfying the transverse and traceless condition $\gamma_{ii}=\partial_i\gamma_{ij}=0$.

Under the linear gauge transformation
\beq
x^\mu\to x^\mu+\xi^\mu\left(t,\textbf{x}\right),
\eeq
these four Stueckelberg fields transform accordingly,
\beq
\pi^0\to\pi^0+\dot{\varphi}\xi^0,\qquad \pi^a\to\pi^a+\xi^i\delta_i^a.
\eeq
On the other hand, the  vector $Z^{\mu}$ defined by \cite{Gumrukcuoglu:2011zh}
\beq
Z^0\equiv\frac{-a}{N}\beta+\frac{a^2}{2N^2}\dot{E},\qquad Z^i\equiv\frac{1}{2}\delta^{ij}\left(\partial_jE+F_{j}\right),
\eeq
transforms in the same manner 
\beq
Z^\mu\to Z^\mu+\xi^\mu.
\eeq
Therefore,  the combination $Z^i-\pi^i$ and $\dot{\varphi} Z^0-\pi^0$ are gauge invariant. 
It is very convenient  to transform to the unitary gauge by simply muting all Goldstone bosons, while keeping all perturbation variables in the eq. (\ref{perturbation}). It can be achieved by adopting a proper diffeomorphism $\xi^\mu(t,\textbf{x})$. In this gauge, the Goldstone bosons are eaten by the graviton and the graviton develops a mass gap at the low energy spectrum. The graviton turns into a massive spin-2 particle with at most 5 polarizations (it can be less), if the Ostrogradsky ghost had been eliminated (which will be done soon!).

On the other hand, we can also transform back to the Goldstone bosons' gauge by muting the vector $Z^\mu$, which can also be done by adopting a proper diffeomorphism $\xi^\mu$. Noted that the way back to the Goldstone bosons' gauge is not unique, we may choose to mute some other perturbation variables in the eq. (\ref{perturbation}), based on their transformations under the diffeomprhism. 

\subsection{Hamiltonian analysis in the unitary gauge}\label{HamAna1}
The Hamiltonian analysis was introduced by P. Dirac in the 50s and 60s \cite{dirac}, as a way of counting dynamical degrees of freedom, and  quantizing mechanical systems such as gauge theories (for busy/lazy readers see the appendix A in the Ref. \cite{Lin:2014jga} for a digested version of the method).

It is more convenient to perform the Hamiltonian analysis in the unitary gauge where $\pi^0=\pi^i=0$, as all terms introduced by Stueckelberg fields appear only in the potential sector in the action and thus the Legendre transformation can be easily done. In the unitary gauge, the action eq. (\ref{action1}) reduces to (I set $M_p^2=1$ in this subsection)
\beq
S=\int d^4x\sqrt{-g}\left[\frac{1}{2}\mathcal{R}-\frac{c}{3}\Lambda_1(t)g^{ij}\delta_{ij}+\frac{\lambda}{4}\left(\frac{\Lambda_2(t)}{N^2}-\frac{1}{3}g^{ij}\delta_{ij}\right)\right].
\eeq
where $\Lambda_1=\varphi(t)^{-2}$ and $\Lambda_2(t)=\dot{\varphi}(t)^2$, namely $\varphi$ and $\dot{\varphi}$ should not be treated as a canonical variable and its velocity, and instead, they should be treated as two time-dependent functions subject to some constraint equations which will be derived later. The reason is that In the unitary gauge, all 4 Stueckelberg fields are eaten by the graviton, and turn themselves into ADM variables, or some space-time functions, and cease to be the independent dynamical degrees of freedom.

The Hamiltonian can be obtained by performing the Legendre transformation. To this end, I need to adopt the ADM decomposition, 
\beq
ds^2=-N^2dt^2+h_{ij}\left(dx^i+N^idt\right)\left(dx^j+N^jdt\right),
\eeq
the inverse of the metric reads,
\beq
g^{00}=-\frac{1}{N^2},\qquad g^{0i}=\frac{N^i}{N^2},\qquad g^{ij}=h^{ij}-\frac{N^iN^j}{N^2}.
\eeq
The conjugate momenta are defined in the following, 
\beq
\Pi^{ij}=\frac{\partial \mathcal{L}}{\partial \dot{h}_{ij}}=\frac{1}{2}\sqrt{h}\left(K_{ij}-Kh_{ij}\right),\qquad \Pi_N=\frac{\partial\mathcal{L}}{\partial \dot{N}}&=&0,\nonumber\\
\Pi_i=\frac {\partial\mathcal{L}}{\partial\dot{N}^i}=0,\qquad \Pi_\lambda=\frac{\partial\mathcal{L}}{\partial\dot{\lambda}}&=&0.
\eeq
The Hamiltonian is obtained by performing the Legendre transformation,
\beq
H&=&\int d^3x\left(\Pi^{ij}\dot{h}_{ij}-\mathcal{L}+\varrho_N\Pi_N+\varrho^i\Pi_i+\varrho_\lambda\Pi_\lambda\right)\nonumber\\
&=&\int d^3x \sqrt{h}\left(\mathcal{H}+\varrho_N\Pi_N+\varrho^i\Pi_i+\varrho_\lambda\Pi_\lambda\right)
\eeq
where $\varrho_N, \varrho^i, \varrho_\lambda$ are Lagrangian multipliers, and 
\beq
\Pi_N\approx0,\qquad \Pi_i\approx0,\qquad\Pi_\lambda\approx0
\eeq
are 5 primary constraints, and 
\beq\label{ham1}
\mathcal{H}&=&\frac{2N}{h}\left(\Pi^{ij}\Pi_{ij}-\frac{1}{2}\Pi^2\right)-\frac{NR}{2}+N\left(\frac{c}{3}\Lambda_1(t)+\frac{\lambda}{4}\right)\left(h^{ab}-\frac{N^aN^b}{N^2}\right)\delta_{ab}-\frac{\lambda}{4}\frac{\Lambda_2(t)}{N}\nonumber\\
&&-2\nabla_j\left(\frac{\Pi^j_{~i}}{\sqrt{h}}\right)N^i,
\eeq
where $R$ is the 3-dimensional Ricci scalar,  $\nabla_i$ is the covariant derivative compatible with 3-dimensional induced metric $h_{ij}$, and $\approx$ denotes the ``weak equivalence", namely  equalities hold on the constraint surface. Caution should be paid to the terms with different types of indices. For instance, in the 3 dimensional hyperspace $h^{ab}\equiv h^{ij}\delta_i^a\delta_j^b$ and $N^a\equiv N^i\delta_i^a$ are scalars (as their space-time indices are dummy indices), $\delta^a_i$ is a vector, while $h^{ij}$ and $N^i$ are tensor and vector respectively.  The consistency conditions of these 5 primary constraints give rise to the following 5 secondary constraints, 
 \beq
\frac{d\Pi_N}{dt}&=&\{\Pi_N,H\}=-\mathcal{H}_0\approx0,\nonumber\\
\frac{d\Pi_i}{dt}&=&\{\Pi_i,H\}=-\mathcal{H}_i\approx0,\nonumber\\
\frac{d\Pi_\lambda}{dt}&=&\{\Pi_\lambda,H\}=-\mathcal{C}_\lambda\approx0,
 \eeq
 where $\mathcal{H}_0$ is the Hamiltonian constraint, $\mathcal{H}_i$ is the momentum constraint, and $\mathcal{C}_\lambda$ is the constraint introduced by hand at the action level,
 \beq\label{ham2}
 \mathcal{H}_0&\equiv& \frac{2}{h}\left(\Pi^{ij}\Pi_{ij}-\frac{1}{2}\Pi^2\right)-\frac{R}{2}+\left(\frac{c}{3}\Lambda_1(t)+\frac{\lambda}{4}\right)\left(h^{ab}+\frac{N^aN^b}{N^2}\right)\delta_{ab}+\frac{\lambda}{4}\frac{\Lambda_2(t)}{N^2},\nonumber\\
 \mathcal{H}_i&\equiv& -2\nabla_j\left(\frac{\Pi^j_{~i}}{\sqrt{h}}\right)-\frac{2N^b\delta_{ab}\delta_{i}^a}{N}\left(\frac{c}{3}\Lambda_1(t)+\frac{\lambda}{4}\right),\nonumber\\
  \mathcal{C}_\lambda&\equiv&\frac{N}{4}\left(h^{ab}-\frac{N^aN^b}{N^2}\right)\delta_{ab}-\frac{\Lambda_2(t)}{4N}.
 \eeq
These 5 secondary constraints must be conserved in time, which yields to the following 5 consistency conditions, 
 \beq\label{cons2}
 \frac{d\mathcal{H}_0}{dt}=\frac{\partial \mathcal{H}_0}{\partial t}+\{\mathcal{H}_0,H\}&\approx&0,\nonumber\\
\frac{d\mathcal{H}_i}{dt}=\partial_t\mathcal{H}_i+\{\mathcal{H}_i,H\}&\approx&0,\nonumber\\
\frac{d\mathcal{C}_\lambda}{dt}=\partial_t\mathcal{C}_\lambda+\{\mathcal{C}_\lambda,H\}&\approx&0.
 \eeq
Whether these 5 consistency conditions generate some tertiary constraints, or they only fix the Lagrangian multipliers in the Hamiltonian, crucially depends on the rank of the matrix
\beq
\mathcal{M}_{AB}\equiv\{\phi_A,\phi_B\},
\eeq
where $\phi_A$ is the whole set of constraints  that we have so far, i.e. $\phi_A=(\Pi_N,\Pi_i,\Pi_\lambda,\mathcal{H}_0,\mathcal{H}_i,\mathcal{C}_\lambda)$. The commutation relations among these constraints are showed in the following,
 \beq
 \{\Pi_N,\Pi_N\}&\approx&0,\qquad \{\Pi_N,\Pi_i\}\approx0,\qquad\{\Pi_N,\Pi_\lambda\}\approx0,\nonumber\\
 \{\Pi_N,\mathcal{H}_0\}&\neq&0,\qquad\{\Pi_N,\mathcal{H}_i\}\neq0,\qquad\{\Pi_N,\mathcal{C}\}\neq0,\nonumber\\
 \{\Pi_i,\Pi_j\}&\approx&0,\qquad\{\Pi_i,\Pi_\lambda\}\approx0,\qquad\{\Pi_i,\mathcal{H}_0\}\neq0,\nonumber\\
 \{\Pi_i,\mathcal{H}_j\}&\neq&0,\qquad\{\Pi_i,\mathcal{C}_\lambda\}\neq0,\nonumber\\
 \{\Pi_\lambda,\Pi_\lambda\}&\approx&0,\qquad\{\Pi_\lambda,\mathcal{H}_0\}\neq0,\qquad\{\Pi_\lambda,\mathcal{H}_i\}\approx0,\nonumber\\
 \{\Pi_\lambda,\mathcal{C}_\lambda\}&\approx&0,\nonumber\\
 \{\mathcal{H}_0,\mathcal{H}_0\}&\neq&0,\qquad\{\mathcal{H}_0,\mathcal{H}_i\}\neq0,\qquad\{\mathcal{H}_0,\mathcal{C}_\lambda\}\neq0,\nonumber\\
 \{\mathcal{H}_i,\mathcal{H}_j\}&\neq&0,\qquad\{\mathcal{H}_i,\mathcal{C}_\lambda\}\neq0,\nonumber\\
 \{\mathcal{C}_\lambda,\mathcal{C}_\lambda\}&\approx&0.
 \eeq
The rank of the matrix $\mathcal{M}_{AB}$ is 10. Therefore the consistency conditions eq. (\ref{cons2}) only fix the Lagrangian multipliers $\varrho'$s, instead of generating new tertiary constraints.   Let's collect all of primary and secondary constraints in the total Hamiltonian and treat them on the same footing,
 \beq
 H_{tot}=\int d^3x \sqrt{h}\left(\mathcal{H}+\varrho_N\pi_N+\varrho^i\Pi_i+\varrho_\lambda\Pi_\lambda+\varrho^0\mathcal{H}_0+\tilde{\varrho}^i\mathcal{H}_i+\varrho_\lambda\mathcal{C}_\lambda\right)
 \eeq
 where $\left(\varrho_N,\varrho^i,\varrho_\lambda,\varrho^0,\tilde{\varrho}^i,\varrho_\lambda\right)$ are Lagrangian multipliers, and we have absorbed the shift vector into the $\tilde{\varrho}^i$.  The algebra closes here.
 
 The rank of the matrix also implies all of these 10 constraints are second class. This is the consequence of that the gauge symmetries in GR, namely the space-time diffeomorphisms are all broken in the unitary gauge. Now let's count the number of degrees. In the phase space we have got 22 degrees in the first place, where 20 degrees are from  the 10 independent components of $g^{\mu\nu}$ and their conjugate momenta, and  2 degrees are from Lagrangian multiplier $\lambda$ and its conjugate momentum.  On the other hand, those 10 second class constraints $\phi_A$ remove 10 degrees in the phase space, and the leftover is 
 \beq
 22-10=12 ~\text{ degrees in the phase space},
 \eeq
which corresponds to 6 degrees of freedom in the physical space-time. The sixth mode is the Ostrogradsky ghost. 
 
\subsection{Ghost elimination in the Goldstone action in the decoupling limit}
The Goldstone action in the decoupling limit is a very convenient perspective for us to hunt and eventually execute the ghost. Taking a closer look at the Goldstone action eq. (\ref{goldstone1}), I notice that the ghost arises from the kinetic term of the Goldsteon boson $\pi$, and the constraint  equating the temporal derivative of $\pi^0$ and the gradient of the $\pi$, namely
\beq
\int \partial_i\dot{\pi}\partial_i\dot{\pi}~\to~ \int\ddot{\pi}^0\frac{1}{\partial^2}\ddot{\pi}^0\qquad \text{as}\qquad\partial^2\pi=3\dot{\pi}^0.
\eeq
Therefore, a direct way to eliminate the higher order temporal derivative term is to eliminate the kinetic term $\partial_i\dot{\pi}\partial_i\dot{\pi}$, which can be achieved by introducing the symmetry \cite{Dubovsky:2004sg}
\beq\label{ressym}
\pi^i(t,\textbf{x})\to \pi^i(t,\textbf{x})+\xi^i\left(t\right),
\eeq
where $\xi^i$ is an arbitrary function of time. This symmetry prohibits all temporal derivative terms of the Goldstone pion $\pi^i$, including the ones of the longitudinal mode $\pi$ and the ones of the transverse mode $\hat{\pi}^i$.  At the lowest dimensional operator level, the building block which respects this symmetry is the following, 
\beq\label{zab}
Z^{ab}\equiv\partial_{\mu}\phi^{a}\partial^{\mu}\phi^{b}-\frac{\left(\partial_{\mu}\varphi\partial^{\mu}\phi^{a}\right)\left(\partial_{\nu}\varphi\partial^{\nu}\phi^{b}\right)}{\partial_{\mu}\varphi\partial^{\mu}\varphi}.
\eeq
In the unitary gauge where all Goldstone bosons are muted, we have $Z^{ab}=\frac{1}{3}h^{ij}\delta_{i}^a\delta_{j}^b$, namely the residual symmetry eq. (\ref{ressym}) strips the shift off the $g^{ij}$, and leaves us with only 3 dimensional induced metric $h^{ij}$. 
Now let's look at the leftover action, which reads
\beq
S\simeq\Lambda^4\int -c\dot{\pi}^0\dot{\pi}^0-\frac{1}{3}\left(\lambda+c\right)\partial_i\pi^0\partial_i\pi^0,
\eeq
where $c$ and $\lambda$ are assumed to be positive, to ensure that the energy density of  dark energy and dark radiation are positive. Therefore, the kinetic term of the leftover scalar mode $\pi^0$ has a wrong sign. Unfortunately after removing the problematic higher order temporal derivative term, the theory still contains a ghost. This ghost arises from the gradient term of the Goldstone boson $\pi$, namely 
\beq
\int-\frac{4c+\lambda}{12}\partial^2\pi\partial^2\pi~\to~\int-\left(3c+\frac{3\lambda}{4}\right)\dot{\pi}^0\dot{\pi}^0\qquad \text{as}\qquad \partial^2\pi=3\dot{\pi}^0.
\eeq
I have to flip the sign of the term $\partial^2\pi\partial^2\pi$ to cure this ghost pathology. The remedy is the  operator introduced in Ref. \cite{Lin:2015cqa},
\beq
 \bar{\delta}Z^{ab}\equiv Z^{ab}-3\frac{Z^{ac}Z^{db}\delta_{cd}}{Z^{cd}\delta_{cd}}, 
\eeq
which is traceless up to the linear perturbation level, and thus it does not contribute to the background evolution. Additionally I shall stick to  the  $Z_2$ symmetry and the global scaling invariance eq. (\ref{globalsym}) to tighten up the structure and reduce the arbitrariness of the theory.
At the lowest dimensional operator level, I write down the general covariant and ghost free action which respects the residual symmetry eq. (\ref{ressym}), the global symmetry eq. (\ref{globalsym}) and the $SO(3)$ rotational symmetry in the field space as follows,
\beq\label{action2}
S=&&\int d^4x\sqrt{-g}\left\{\frac{M_p^2}{2}\mathcal{ R}-M_p^2\varphi^{-2}\left[\left(c+\lambda\right)\cdot Z+\lambda\partial^\mu\varphi\partial_\mu\varphi+\frac{3d}{8Z} \cdot\bar{\delta}Z^{ab} \bar{\delta}Z^{cd}\delta_{ac}\delta_{bd}\right]\right\}\nonumber\\
&&\qquad\qquad+S_m\,,
\eeq
where $Z\equiv Z^{ab}\delta_{ab}$, the coefficient $3/8$ is chosen for the later convenience and $S_m$ is the action of the matter sector which minimally couples to gravity. 
This is the main result of my current work. Compared with the original action eq. (\ref{action1}),  in the unitary gauge the $g^{ab}$, which is produced by the spatial Stueckelberg fields $\partial^\mu\phi^a\partial_\mu\phi^b\delta_{ab}$, is replaced by one of our building blocks $h^{ab}$ defined in the eq. (\ref{zab}). Take the note of that $h^{ab}$ differs from $g^{ab}$ only at the perturbation level. On the other hand, the operator $ \bar{\delta}Z^{ab}$ only appears at the perturbation level too as it is traceless. Therefore, all modifications are operated at the perturbation level, and do not alter the background evolution which  is subject to the set of equations in the eq. (\ref{friedmann}) (\ref{bgcons}). The Goldstone action for the scalar graviton reads
\beq
S\supset\int-\left(c+d\right)\dot{\pi}^0\dot{\pi}^0-\frac{1}{3}\left(\lambda+c\right)\partial_i\pi^0\partial_i\pi^0,
\eeq
with the ghost free condition, 
\beq\label{gf1}
c+d<0.
\eeq
This ghost free condition will be reproduced later in the full perturbation analysis in the unitary gauge, where the Goldstone bosons are muted. 

\subsection{Full perturbation analysis in the unitary gauge}

We adopt the metric perturbation decomposition in the eq. (\ref{perturbation}). Due to the background $SO(3)$ rotational invariance, the scalar perturbations, the vector perturbations and the tensor perturbations completely decouple from each other at the linear perturbation level. We also adopt the unitary gauge and mute all Goldstone bosons. 

The massive gravity is an analog of the Higgs mechanism in the particle physics, where the Higgs field resides in an asymmetric state.  In the unitary gauge, the Goldstone pions in this symmetry broken phase are eaten by the gauge bosons and consequently the gauge bosons become massive. We expect the same phenomena should also occur in the massive gravity, namely once we mute the Goldstone bosons, the scalar graviton and gravitational waves are massive in the unitary gauge. 

I will only focus on the perturbation analysis of the pure gravity, while the cases with matter included will be discussed in the future work. 

\subsubsection{Scalar perturbation}
The quadratic action of the scalar perturbation is obtained after a straightforward computation, 
\beq
S_{\text{scalar}}^{(2)}=\int d^4x\left(\mathcal{L}_{\text{EH}}+\mathcal{L}_{\text{mass}}\right),
\eeq
where $\mathcal{L}_{\text{EH}}$ is the contribution from the Einstein-Hilbert action,
\beq
\frac{\mathcal{L}_{EH}}{M_p^2a^3}&=&-3\dot{\psi}^2+\frac{k^2}{a^2}\left[\psi^2-\frac{1}{2}a^2\dot{E}\left(H\psi-2\dot{\psi}\right)-\frac{1}{2}a^2HE\left(\dot{\psi}+3H\psi\right)\right]\nonumber\\
&~~&+\alpha\left[\frac{2k^2\beta H}{a}+\frac{k^2}{a^2}\left(2\psi-a^2H\dot{E}\right)+6H\dot{\psi}-3H^2\alpha\right]-\frac{2k^2\beta\dot{\psi}}{a},
\eeq
and the $\mathcal{L}_{\text{mass}}$ is the contribution from the dark sector, or in other word the graviton mass term, 
\beq
\mathcal{L}_{\text{mass}}&=&M_p^2\varphi^{-2}\left[-\frac{3c+d}{36}k^4aE^2-\frac{1}{6}ck^2aE\left(2\alpha+\psi\right)+ca\psi\left(2\alpha+\psi\right)\right]\nonumber\\
&~&+\lambda\left[-\frac{k^4E^2}{48a}+\frac{k^2E\left(\psi-\alpha\right)}{12a}+\frac{\left(\alpha+\psi\right)^2}{4a}\right]+\delta\lambda\left(\frac{\psi-\alpha}{2a}-\frac{k^2E}{12a}\right).
\eeq
I  have absorbed the lapse into the redefinition of time, and  rescaled the Lagrangian multiplier $\lambda\to\lambda\cdot\frac{\varphi^{2}}{4a^2M_p^2}$ for convenience once again.
Taking the variation of the action with respect to the non-dynamical variables $\alpha$, $\delta\lambda,$ and $~\beta$, I get the following three constraint equations, 
\beq
H\left(6\dot{\psi}-k^2\dot{E}\right)+\frac{2c\psi}{a^2\varphi^2}+\frac{k^2}{a^2}\left(2\psi-\frac{cE}{3a^2\varphi^2}\right)+\frac{\lambda}{12a^4M_p^2}\left(6\psi-k^2E\right)&&\nonumber\\
-\frac{\delta\lambda}{2a^4M_p^2}+\alpha\left(\frac{\lambda}{2a^4M_p^2}-6H^2\right)+\frac{2k^2H\beta}{a}&=&0,\nonumber\\
k^2E+6\textcolor{red}{\alpha}-6\psi&=&0,\nonumber\\
H\alpha-\dot{\psi}&=&0.
\eeq
Substituting the solution of the above equations into the action, the quadratic action of the scalar perturbation takes the following form,
\beq
S_{\text{scalar}}^{(2)}=\int -\frac{\left(c+d\right)aM_p^2}{H^2\varphi^2}\dot{\psi}^2+...
\eeq
where the eclipse denotes the potential term and gradient term. The ghost free condition for the scalar perturbation reads
\beq
c+d<0.
\eeq
We have reproduced the ghost free condition eq. (\ref{gf1}), in the different gauge.
Let me redefine a new constant $b$ via 
\beq
b\equiv -2\left(c+d\right), \qquad \text{and we demand}\qquad b>0.
\eeq
The scalar graviton is canonically normalised  as 
\beq
\psi^c\equiv\frac{\sqrt{b}M_p}{aH\varphi}\psi.
\eeq
The canonically normalised scalar action reads
\beq
S_{\text{scalar}}^{(2)}=\frac{1}{2}\int dtd^3k a^3\left(\dot{\psi}^c\dot{\psi}^c-\frac{c_s^2k^2}{a^2}\psi^c\psi^c-M^2_{s}\psi^c\psi^c\right),
\eeq
where the sound speed reads
\beq
c_s^2=\frac{c}{b}\cdot\frac{\rho_{\text{dark}}}{\rho_{\text{hde}}}\cdot\left(1+w_{\text{dark}}\right),
\eeq
where $\rho_{\text{dark}}$ is energy density of the dark sector defined in the eq. (\ref{rhodark}),  $w_{\text{dark}}$ is its equation of state, and $\rho_{hde}=cM_p^2/a^2\varphi^2$ is the energy density of the holographic dark energy.  It is clear that the $c_s^2$ becomes negative when the equation of state of the dark sector drops below $-1$,  this is precisely the case for $c<6$ in the asymptotic future \cite{Li:2012xf}\cite{Li:2012fj}. Therefore, if we demand the freeness of the gradient instability throughout the whole cosmic expansion history, the parameter region with $c<6$ needs to be excluded. However, $c<6$ is still allowed, if we only demand the freeness of the gradient instability at present and in the past, and allow the instability to occur in the future. The mass of the scalar mode reads 
\beq
M_s^2&=&2H^2\left[\frac{6c}{b}-\Omega_{\text{hde}}^2+\Omega_{\text{hde}}\left(-\frac{1}{2}-\frac{3}{c}+\frac{3c}{b}-4\Omega_{\text{rad}}\right)+\sqrt{\frac{3\Omega_{\text{hde}}}{c}}\left(1-\frac{2c}{b}+2\Omega_{\text{rad}}\right)\right.\nonumber\\
&~&\qquad\left.+\left(1+\frac{8c}{b}\right)\Omega_{\text{rad}}+\Omega_{\text{rad}}^2\left(\frac{4c}{b\Omega_{\text{hde}}}-4\right) \right],
\eeq
where
 $\Omega_{rad}$ is the fraction of the dark radiation defined by $\Omega_{rad}\equiv\rho_{rad}/3M_p^2H^2$, and $\Omega_{\text{hde}}$ is the fraction of the holographic dark energy defined by $\Omega_{\text{hde}}\equiv \rho_{\text{hde}}/3M_p^2H^2$. 
The Compton wavelength of the scalar mode is about Hubble radius size, up to a coefficient whose value is dependent on the parameters $b$ and $c$.

One may wonder whether the mass of scalar graviton is always positive along the cosmic evolution to ensure the stability in the scalar sector. This problem may require a thorough numerical analysis, and thus it will not be carried out in the current work. Instead, I would like to  show the tachyon freeness condition  in the asymptotic de-Sitter phase where $c=6$ \cite{Li:2012xf} and  $w_{\text{dark}}\to-1$. The scalar mass in the asymptotic phase reads
\beq
M_s^2\to 8H^2\left[\frac{6}{b}-1\right].
\eeq
The tachyon freeness condition translates to 
\beq\label{stcod}
0<b<6.
\eeq
The stability analysis for general space-time solution will be performed in the future work. 
\subsubsection{Tensor perturbation}
The quadratic action for the tensor perturbation reads
\beq
S_{\text{tensor}}^{(2)}=\frac{M_p^2}{8}\int dtd^3ka^3\left[\dot{\gamma}^{ij}\dot{\gamma}_{ij}-\left(\frac{k^2}{a^2}+M^2_{GW}\right)\gamma_{ij}\gamma^{ij}\right],
\eeq
where 
\beq
M_{GW}^2=
\frac{1}{6R_h^{2}}\left[4c-b+3c\frac{\rho_{\text{dark}}}{\rho_{\text{hde}}}\left(1+w_{dark}\right)\right]
\eeq
The Compton wavelength of the tensor graviton is about the size of the future event horizon.  The following tachyon freeness condition is required
\beq
4c-b+3c\frac{\rho_{\text{dark}}}{\rho_{\text{hde}}}\left(1+w_{dark}\right)>0
\eeq
to ensure the stability of the tensor sector.  In the asymptotic de-Sitter phase where $c=6,~w_{\text{dark}}=-1$, the stability condition translates to $0<b<24$, which is weaker than the stability condition eq. (\ref{stcod}) derived in the preceding subsection of the scalar mode analysis. 

The non-vanishing mass modifies the dispersion relation of the gravitational waves, and the group velocity is constrained with high precision $-3\cdot 10^{-15}<c_{\text{gw}}-1<7\cdot10^{-16}$ at low redshift regime by the multi-messenger observation GW170817 \cite{Monitor:2017mdv}. It gives us an upper bound on the graviton mass around $m_g<10^{-23}\text{eV}$. Our graviton mass is much lower than this upper bound by around 10 orders of magnitude. It remains  challenging to directly probe the non-vanishing graviton mass effects at the late time epoch. However, the size of the future event horizon is much shorter during the early universe, and thus it leads to a much larger graviton mass, which may give rise to some interesting observational effects on the stochastic gravitational waves background.

\subsubsection{Vector perturbation}
I do not expect to find the dynamical degrees of freedom in the vector sector at the linear perturbation level, as the residual symmetry eq. (\ref{ressym}) has projected out all temporal derivative of the Goldstone vector bosons. I will try to confirm it in this subsection. The quadratic action of the vector perturbation reads
\beq
S_{\text{V}}^{(2)}=\frac{M_p^2}{16}\int dtd^3kk^2a^3\left[\dot{F}_i\dot{F}^i-\frac{4S_i\dot{F}^i}{a}+\frac{4S_iS^i}{a^2}+\left(\frac{b-6c}{6a^2\varphi^2}-\frac{\lambda}{3M_p^2a^4}\right)F_iF^i\right].
\eeq
Taking the variation with respect to the Lagrangian multiplier $S_i$, I get 
\beq
a\dot{F}_i-2S_i=0.
\eeq
Substituting the solution of this constraint equation back to the action, I get 
\beq
S_{\text{V}}^{(2)}=\frac{1}{16}M_p^2\int dtd^3kk^2a^3\left(\frac{b-6c}{6a^2\varphi}-\frac{\lambda}{3M_p^2a^4}\right)F_iF^i.
\eeq
The kinetic term of the vector mode has vanished, and the action is subject to a new constraint $F_i=0$. After substituting this solution back to the action, the whole action vanishes and I conclude that no dynamical degree of freedom is found in the vector sector at the linear perturbation level. Whether there exists dynamical degrees of freedom at the higher order level requires a fully non-linear analysis, which will be provided in the next subsection. 

\subsection{The Hamiltonian analysis confirms the total number of degrees}
I have performed the Hamiltonian analysis in the subsection (\ref{HamAna1}), and I found that the original theory contains 6 modes. Compared to the Hamiltonian of the original theory (\ref{action1}), the modified one (\ref{action2}) strips off the term that is non-linear in the shift vector. Consequently the primary constraint $\Pi_i\approx0$ commutes with all primary and secondary constraints. The consistency conditions for the momentum constraints conserve in time yield to another 3 tertiary constraints,
\beq
\mathcal{T}_i\equiv\{\mathcal{H}_i,H\}\approx0.
\eeq
Now let's add these 3 new tertiary constraints into the constraint set $\phi^A$, and remove the $\Pi_i\approx0$ as it is first class, we have the new set
\beq\label{conset2}
\phi_A=(\Pi_N,\Pi_\lambda,\mathcal{H}_0,\mathcal{H}_i,\mathcal{C}_\lambda,\mathcal{T}_i).
\eeq
We can show that the rank of the following  $10\times10$ matrix 
\beq
\mathcal{M}_{AB}\equiv\{\phi_A,\phi_B\}
\eeq
is 10, no new constraints are generated and the algebra closes here. Therefore, all  constraints in the eq. (\ref{conset2}) are second class, and they remove 10 degrees in the phase space. On the other hand, the constraints $\Pi_i\approx0$ remove 6 degrees in the phase space as they are first class. The total number of the leftover dynamical degrees of freedom  in the phase space is 
\beq
22-3\times2-10=6,
\eeq
which corresponds to  3 dynamical degrees of freedom in the physical space-time. As we have shown in our perturbation analysis, 2 of them are tensor modes, and the rest  one is a scalar graviton. The  additional degrees of freedom are fewer than the broken symmetry generators, because the Lorentz-invariance of the background space-time is broken and therefore the Goldstone theorem needs to be understood in the extended sense.

\subsection{Physical interpretation of the UV-IR correspondence}
The UV-IR correspondence is based on the assumption that the vacuum energy arising from the quartic divergence should not exceed the energy density of a black hole in the same size, and the effective quantum field theory should break down at the Schwarzschild radius scale. Take the note of that the Hubble radius  of our universe coincides with its Schwarzschild radius, and thus the bound should be saturated on the cosmological background and we have the UV-IR relation $\Lambda\sim\sqrt{M_p/L}$.

One may ask how this UV-IR correspondence appears naturally in our effective field theory framework? The answer is that it should be interpreted as the scale where the scalar graviton strongly couples to itself. Let's look at the 4-votex interaction in the decoupling limit, 
\beq
\int M_p^2\varphi^{-2}\partial^2\pi\partial^2\pi\partial^2\pi\partial^2\pi\sim\int M_p^2\varphi^{-2}\dot{\pi}^0\dot{\pi}^0\dot{\pi}^0\dot{\pi}^0\sim\int \frac{\omega^4}{M_p^2\varphi^{-2}}\pi^0_c\pi^0_c\pi^0_c\pi^0_c,
\eeq
where $\pi^0_c$ is the canonical normalised Goldstone boson, and $\omega$ is the frequency of $\pi^0_c$. The strength of the coupling exceeds unity if $\omega> \sqrt{M_p/\varphi}$, and our effective field theory breaks down. As shown in our perturbation analysis, we have learnt that $\varphi$ is the graviton's Compton's wavelength (remember $\varphi$ has the length dimension, and we set the scale factor $a=1$ in the decoupling limit). Therefore, a  sensible and physical interpretation of this UV-IR correspondence is that the UV cut-off scale of our EFT is inversely proportional to the Compton wavelength of our graviton. 

\section{Conclusion and Discussion}\label{sec4}
In the current work, a general covariant local field theory  is proposed for the holographic dark energy model. I started from the mini superspace action proposed a few years ago  \cite{Li:2012xf}, showed that the direct, and perhaps the simplest way of covariantizing the action gives rise to additional 4 degrees of freedom (6 in total including 2 degrees in the gravitational wave sector), which leads to the Ostrogradsky ghost instability. To remedy the ghost pathology, I introduced a new symmetry to prohibit the problematic terms in the theory. It has turned out that the low energy effective field theory of the holographic dark energy model is actually the Lorentz-violating massive gravity theory, whose graviton has 3 polarisations including 1 helicity 0 mode and 2 helicity 2 modes. The Compton wavelength of the graviton is about the size of the future event horizon of our universe, as shown in the linear perturbation analysis in the unitary gauge. To confirm the total number of dynamical degrees of freedom of our theory at the fully non-linear level, I have performed the Hamiltonian analysis and found that the momentum constraints yield to another 3 tertiary constraints. These 6 constraints, which are all second class, eliminate 3  out of 4 additional degrees and therefore there are only 3 dynamical degrees of freedom in total in the gravity sector, including a scalar mode and two tensor modes. Our effective field theory breaks down at the scale $\Lambda\sim\sqrt{M_p/L}$, where $L$ is the graviton's Compton wavelength, due to the strong coupling of the scalar graviton above this scale, which offers a natural and physical interpretation for the UV-IR correspondence in the holographic dark energy model.

Our effective field theory provides a general framework in which the holographic dark energy model can be tested. For instance, the non-vanishing mass  of the gravitational waves, which is small at late time epoch but sizeable during the early universe,  leads to a modified stochastic gravitational waves background. On the other hand, the existence of the scalar graviton whose Compton wavelength is about the size of the Hubble radius (up to  an order $\mathcal{O}(1)\sim\mathcal{O}(10)$ coefficient) during inflation may be tested by the cosmological collider physics \cite{ccp}. It is also very intriguing to ask what would happen if the Stueckelberg field, say the time like one $\varphi$, couples to the standard model fields. Our EFT framework also provides a general setup where the perturbation theory of the holographic dark energy can be developed, which allows us to investigate its impacts on the cosmological structure formation in detail.  All these possibilities warrant further scrutiny. 

I shall end by commenting that the global Lorentz invariance is broken in our scalar field configuration. As an analog to the Higgs mechanism in the particle physics, one may expect a similar symmetry restoration to occur at high energy scale, where the Lorentz invariance is recovered and the graviton becomes massless again. However, this Higgs-like mechanism is till missing. In fact, to my best knowledge it is still an open question for all massive gravity theories. 

\section*{Acknowledgement}
This work is supported by the grant No. UMO-2018/30/Q/ST9/00795 from the National Science Centre, Poland. The author would like to thank Yi-fu Cai, Alexander Ganz, A. Emir Gümrükçüoğlu, Qing-Guo Huang, Miao Li, Yin-zhe Ma, and Yi Wang for the useful discussions and suggestions.

\end{document}